# CRITICAL FIELD OF Al-DOPED MgB$_2$ SAMPLES: CORRELATION WITH THE SUPPRESSION OF σ-BAND GAP

M.Putti[1], C.Ferdeghini[1], M.Monni[1], I.Pallecchi[1], C.Tarantini[1], P.Manfrinetti[2], A.Palenzona[2], D.Daghero[3], R.S.Gonnelli[3], V.A. Stepanov[4]

[1] INFM-LAMIA, Dipartimento di Fisica, Università di Genova, via Dodecaneso 33, 16146 Genoa, Italy
[1] INFM-LAMIA, Dipartimento di Chimica e Chimica Industriale, Università di Genova, via Dodecaneso 33, 16146 Genoa, Italy
[3] INFM-LAMIA, Dipartimento di Fisica, Politecnico di Torino, corso Duca degli Abruzzi 24, 10129 Torino, Italy
[4] P.N. Lebedev Physical Institute, Russian Academy of Sciences, Leninski Prospect 53, 119991, Moscow, Russia

In this Letter, the study of the effect of Al substitution on the upper critical field, $B_{c2}$, in Al$_x$Mg$_{1-x}$B$_2$ samples is presented. We find a straightforward correlation between $B_{c2}$ and the σ–band gap, $\Delta_\sigma$, evaluated by point-contact measurements. Up to x=0.2 $B_{c2}$ can be well described within a clean limit model and its decrease with x is directly related to the suppression of $\Delta_\sigma$. For larger doping we observed the crossover to the dirty regime driven mostly by the strong decrease of $\Delta_\sigma$ rather than by the increase of the σ-band scattering rate.

Superconductivity at 40K in magnesium diboride has been extensively studied since its discovery[1] to these days. The presence of two-bands crossing the Fermi level and having strongly different character[2,3] is well established: two π bands are formed by $p_z$ orbitals of boron and are three-dimensional, electron-type and weakly coupled with phonons; two σ bands are formed by $sp^2$-hybrids orbitals stretched along boron-boron bonds and are two-dimensional, hole-type and strongly coupled with the optical E$_{2g}$ phonon mode. This peculiar band structure, joined to the fact that impurity interband scattering is inhibited by the different parity of π and σ orbitals,[4] is at the ground of the two-gap superconductivity, evidenced unambiguously for the first time in this compound. The lack of interband scattering yields important consequences in the transport behavior because it prevents the mixing of the σ and π carriers, which maintain their own characteristics. This gives a unique chance to selectively disorder each channel independently and offers many opportunities to tune superconducting and normal properties.

Many attempts of selective doping have been carried out. For example, substitution of C in the B site significantly increases upper critical fields, as reported by several groups[5-7] suggesting that, in these substituted compounds, the dirty regime is well stabilized. On the other hand, Al substitution of Mg does not give univocal results. In Al doped single crystals, the critical field perpendicular to the ab-planes, $B_{c2\perp ab}$, increases with increasing Al concentration, while the one parallel to the ab- planes, $B_{c2//ab}$, decreases, as observed also in polycrystalline samples.[9] The behavior of $B_{c2\perp ab}$ has been explained within a dirty limit model, but this interpretation clashes with the decreasing of $B_{c2//ab}$. Actually, there are several effects caused by the Al substitution of Mg: the rising of the Fermi level,[10,11] the stiffening of the E$_{2g}$ mode[12] and the increasing of the interband scattering.[13] The importance of the last effect on the superconducting properties of Al doped samples has not been clarified yet. Its main signature should be the decrease of the σ–band gap, $\Delta_\sigma$, together with the increase of the π−band gap, $\Delta_\pi$, and for large interband scattering the gaps should merge to the BCS value. The only available data of gaps as function of Al[15] show a strong decrease of $\Delta_\sigma$, which seems to be related to the increased interband scattering, but the $\Delta_\pi$ remains rather constant in contrast with expectations. Therefore both the energy gaps, and the scattering rates, vary with the doping determining the regime of conduction (clean or dirty) of each band. Thus the upper critical field must depend on Al doping in a not trivial way and the study of its behavior gives an unique opportunity to investigate the peculiar role of disorder in a two-gap superconductor.

This Letter tackles the multifaced subject of transport regimes in Al doped samples. We correlated the upper critical fields, the energy gaps and the scattering rates in a complete set of poly-crystalline Al$_x$Mg$_{1-x}$B$_2$ samples in order to investigate the effect of Al substitution on the conduction regimes of σ and π bands.

Dense, clean and hard cylinder shaped samples have been prepared by direct synthesis of pure elements.[16] In order to improve the homogeneity of doping, Mg-Al alloys were prepared in a first step with Al concentration ranging from 0 to 0.3. In a second step Mg-Al alloy and crystalline B put in Ta crucibles welded in argon and closed in quartz tubes under vacuum, were heated up to 950 °C. X-ray powder patterns were obtained by a Guinier-Stoe camera. No extra peaks due to the presence of free Mg or spurious phases were detected. Peaks of the doped compounds are shifted in comparison with pure MgB$_2$ peaks and they broaden on increasing the Al content; the lattice parameters are reported in table I and they are in good agreement with other reports.[15,17]

All the samples were cut in the shape of a parallelepiped bar (~1×2×12 mm$^3$). Magnetoresistivity measurements were performed from 0 to 9 T in a PPMS Quantum Design system. The upper critical field curve $B_{c2}(T)$ was operatively defined at 90% of the resistive transition; in polycrystalline samples, this criterion probes a percolation path of grains with the largest critical fields, having their ab crystalline plane parallel to the applied magnetic field.[5]

Point-contact measurements were carried out by using the "soft" technique described elsewhere,[18] that consists in creating a small (nearly 50 μm of diameter) metallic contact on the sample surface by using a drop of Ag. Thanks to these very stable contacts, the resulting low-temperature conductance curves (dI/dV vs. V) presented clear structures due to Andreev reflection at the interface, in the form of conductance maxima whose positions give a rough indication of the gaps amplitude.

In table I, the critical temperature $T_c$, the residual resistivity $\rho(40)$ and the residual resistivity ratio (RRR) of the seven samples are reported. The transition width gets larger for more heavily doped samples, where inhomogeneities are more likely to occur. The resistivity values are comparable with the ones reported for Al doped single crystals[8] indicating the excellent quality of this set of samples. In general, the residual resistivity $\rho_0$ monotonically increases with increasing doping, ranging from 2.5 $\mu\Omega$·cm for the undoped sample to 24 $\mu\Omega$·cm for the x=0.3 sample.

Increasing disorder in single-gap superconductors is usually accompanied by an increase in the upper critical field.[19] In the right panel of figure 1 $B_{c2}$ is plotted as a function of temperature for all the samples considered here. As clearly seen, $B_{c2}$ monotonically decreases with increasing Al concentration; this monotonic trend is maintained even if $B_{c2}$ is plotted as a function of the reduced temperature $T/T_c$, as seen in the left panel of figure 1. Measurements on Al doped single crystals show indeed that the critical field decreases with doping when the external field is parallel to the *ab* planes.[8] In ref. 9 we showed that at low level of doping (x<0.1) this unusual behavior is explained assuming that, in spite of the disorder introduced by Al doping, the $B_{c2}$ can be described within a clean limit model.

In clean limit the critical fields of MgB$_2$ at low temperature are determined by the $\sigma$ bands.[20,21] Thus, far from the transition, the critical fields perpendicular and parallel to the *ab*-planes can be expressed in terms of the coherence lengths of the $\sigma$ carriers in either directions, $\xi_{0\sigma ab}$ and $\xi_{0\sigma c}$:

$$B_{c2\perp ab}(T) = \frac{\Phi_0}{2 \cdot \pi \cdot \xi_{0\sigma ab}(T)^2} \quad (1a)$$

$$B_{c2//ab}(T) = \frac{\Phi_0}{2 \cdot \pi \cdot \xi_{0\sigma ab}(T) \cdot \xi_{0\sigma c}(T)} \quad (1b)$$

By inserting $\xi_{0\sigma j} = \frac{\hbar \cdot v_{F_{\sigma j}}}{\pi \cdot \Delta_\sigma}$ and $\Phi_0 = 2.07 \cdot 10^{-15}$ T·m$^2$ we can express the upper critical fields and their anisotropy as a functions of $\Delta_\sigma$ and of the in-plane, $v_{F_{\sigma ab}}$, and out-of-plane, $v_{F_{\sigma c}}$, Fermi velocities of $\sigma$ bands:

$$B_{c2\perp ab}(T) = \frac{\Phi_0 \cdot \pi \cdot \Delta_\sigma(T)^2}{2 \cdot \hbar^2 \cdot (v_{F_{\sigma ab}})^2} \quad (2a)$$

$$B_{c2//ab}(T) = \frac{\Phi_0 \cdot \pi \cdot \Delta_\sigma(T)^2}{2 \cdot \hbar^2 \cdot v_{F_{\sigma ab}} \cdot v_{F_{\sigma c}}} \quad (2b)$$

$$B_{c2\perp ab} / B_{c2\perp c} = v_{F_{\sigma ab}} / v_{F_{\sigma c}} \quad (2c)$$

Therefore to analyze the upper critical fields, it is necessary to consider the changes in the electronic structure and energy gaps due to Al doping.

By doping with Al the topology of $\sigma$ Fermi surface changes: at x=0.33 (close to the maximum Al concentration we considered) the complete filling of the $\sigma$ bands at the $\Gamma$ point occurs and a crossover happens from a two-dimensional to a three-dimensional dispersion regime.[10,11]

The Fermi velocities of the $\sigma$ bands, defined by

$$v_{F_{\sigma i}}^2 = \left\langle \left( \frac{\partial \varepsilon(\vec{k})}{\partial k_i} \right)^2 \right\rangle_{FS,\sigma}$$

($i=ab, c$) have been calculated by averaging squared velocities over the Fermi surface sheet of the $\sigma$ bands. We used the electronic structure $\varepsilon(\vec{k})$ at various Al content calculated *ab initio* by G. Profeta et al..[11] The Fermi velocities vary with Al doping as shown in the upper panel of figure 2: $v_{F_{\sigma ab}}$ decreases with increasing x while $v_{F_{\sigma c}}$ increases. Thus the critical fields are affected in different ways by the changes in the Fermi velocities: $B_{c2//ab}$ is only slightly affected, while $B_{c2\perp ab}$ has to increase and their anisotropy $\gamma$ (eq. (2c)) has to decrease ($\gamma \approx 6$ at x=0 and $\gamma \approx 4$ at x=0.3). Noticeably, an increase of $B_{c2\perp ab}$ as well as a decrease of $\gamma$ with Al doping have been recently reported in single crystals.[8]

Eq.s (2a) and (2b) show a quadratic dependence on $\Delta_\sigma$, thus, the $\sigma$ gap determination as a function of doping becomes essential. Here, the gap energies have been evaluated by point-contact measurements. A typical low-temperature conductance curve is reported in the inset of Fig. 3. Notice that clear peaks related to $\Delta_\sigma$ are present and, because of thermal smearing, the maxima due to $\Delta_\pi$ have merged in a single peak at zero bias. The solid line superimposed to the experimental data is a fit with the Blonder-Tinkham-Klapwijk (BTK) model generalised to the two-band case[18,22] that gives an accurate evaluation of the gap amplitudes. In fig 3 we plot the so determined $\Delta_\sigma$ and $\Delta_\pi$ in a selected set of samples ($x$ =0, 0.1, 0.2 and 0.3). Because of the dominant contribution of the $\pi$ bands to the conductance across the junction even in polycrystalline samples,[23,24] the uncertainty on the $\Delta_\sigma$ is generally larger than that on $\Delta_\pi$, as evidenced by the data summarized in table II. Note that $\Delta_\sigma$ linearly decrease with Al doping from 7.5 meV at x=0 to 2 meV at x =0.3; $\Delta_\pi$ (2.8 meV at x=0) shows a decrease in the whole doping range becoming as small as 0.5 meV at x=0.3. In the same figure we compare the results obtained by specific-heat measurements.[15] Apart from the absolute values of the gaps, the overall agreement between the data as function of doping is impressive considering differences inherent to the two experimental technique (heat capacity measures an average gap throughout the sample, while Andreev reflection probes the gap in a local area close to the surface). Thus point-contact measurements in Al doped samples confirm the previous evidences of the strong decrease of $\Delta_\sigma$, not accompanied by the rising of $\Delta_\pi$. Moreover we note that for x=0.3 the reduced $\sigma$ gap $2\Delta_\sigma / k_B T_c$ is 2, much smaller than the BCS value. All these outcomes contradict the expectations of the two-bands model and require further clarifications.

Once the Fermi velocities and $\sigma$-band gap have been estimated as functions of Al doping, it should be possible to scale the experimental critical field curves according to eq. (2b) (as previously remarked, our experimental values of $B_{c2}$ must be compared with $B_{c2//ab}$).

In figure 4 we plot the rescaled critical fields, $B_{c2}^{res}(x) = B_{c2}(x) \frac{\Delta_\sigma(0)^2}{\Delta_\sigma(x)^2} \frac{v_{F_{\sigma ab}}(x)}{v_{F_{\sigma ab}}(0)} \frac{v_{F_{\sigma c}}(x)}{v_{F_{\sigma c}}(0)}$, where $B_{c2}(x)$ is the experimental critical field for each x value and $\Delta_\sigma(x)$ is either the measured or the linearly interpolated gap.

Except for the x=0.3 sample, whose curve lies apart far from the others, all the curves fall within the shaded area which represents $B_{c2}(0)$, taking the errors on $\Delta_\sigma$ into account. Actually all the curves fall in the upper side of the shaded region. This can be due to the fact that in not homogeneous samples $B_{c2}$ is determined by the regions with the highest $T_c$ that is with the lowest doping; thus the $\Delta_\sigma(x)$ values could be slightly underestimated and consequently $B_{c2}^{res}$ overestimated. This is true especially for the $Al_{0.3}Mg_{0.7}B_2$ sample which is the least homogeneous. Anyway the overall behavior clearly suggests that, for x≤0.2 a description within a clean limit model works, while a further increase of Al drives abruptly the system in the dirty regime. The crossover from a clean to a dirty regime is not evident from fig. (1) which shows a monotonic decrease of $B_{c2}$ with increasing doping up to x=0.3. But this description is reliable considering that at x=0.3 $\Delta_\sigma$ is decreased down to 2 meV which corresponds to a coherence length as large as 50 nm, probably much larger than the mean free path of σ-band carriers.

Further insights of our analysis comes from the comparison of the σ-band scattering rates, $\Gamma_\sigma$, with the energy gaps $\Delta_\sigma$. An estimation of $\Gamma_\sigma$, can be made by assumunig that the electrical conductivity of Al doped samples is mainly determined by σ-band carriers. In fact being the σ-orbitals localized around the B planes, the σ–band scattering rate is only slightly affected by substitution in the Mg planes.[4] This hypothesis, although reasonable, provides lower limits for $\Gamma_\sigma$ values. However, even if it is relaxed, our conclusions are substantially unchanged. We assume:

$$\rho_{0_\sigma} = \Gamma_\sigma /(\varepsilon_0 \omega_{p_\sigma}^2) \approx \rho_0 \qquad (3)$$

where $\omega_{p_\sigma}$ is the σ-band plasma frequency and $\varepsilon_0$ is the dielectric constant. The plasma frequencies are given by $\omega_{p_{\sigma i}}^2 = 8\pi \frac{N_\sigma}{V_{cell}} v_{F_{\sigma i}}^2$ and can be calculated taking into account the dependence on Al of the Fermi velocities, the density of states $N_\sigma$ and the unit cell volume $V_{cell}$. $\omega_{p_{\sigma ab}}$, $\omega_{p_{\sigma c}}$ and $\omega_{p_\sigma} = \left(\frac{2}{3}\omega_{p_{\sigma ab}}^2 + \frac{1}{3}\omega_{p_{\sigma c}}^2\right)^{1/2}$ are reported in the lower panel of figure 2 as a function of Al doping. As clearly seen, $\omega_{p_\sigma}$ strongly decreases with the doping due to the filling of σ bands. Estimation of $\Gamma_\sigma$ for the various samples can be obtained by equation (3), using the experimental resistivity values and the calculated plasma frequencies. The results are given in table I. The $\Gamma_\sigma$ values weakly increase with Al concentration ranging from 7 to 16 meV. However the increase of resistivity is mostly accounted for by the decrease of $\omega_{p_\sigma}$. Finally in table I the ratios $\frac{\Gamma_\sigma}{\pi\Delta_\sigma} = \frac{\hbar v_{F_\sigma}/\pi\Delta_\sigma}{\hbar v_{F_\sigma}/\Gamma_\sigma} = \frac{\xi_{0\sigma}}{l_\sigma}$ are reported. Except for the case of the x=0.3 sample, the ratios $\Gamma_\sigma/\pi\Delta_\sigma$ turn out to be lower than 1 confirming that the σ bands in Al-doped samples are in clean limit for x≤0.2. For the most substituted sample the transition to the dirty regime evidenced in the critical field analysis, seems to be driven more by the strong decrease of $\Delta_\sigma$ than by the increase of $\Gamma_\sigma$. This view is also consistent with the low $B_{c2}$ presented by this sample.

In conclusion, in this Letter we discussed for the first time the straightforward correlation existing between the critical fields and σ-band gaps in Al doped samples. Our main experimental evidences are that $B_{c2}$ monotonously decreases with Al doping while the resistivity increases. On the other hand, we estimated a suppression of $\Delta_\sigma$ with increasing of Al doping. This complex phenomenology can be understood by assuming that Al substitution in Mg sites while suppressing the conduction of π bands, only slightly affects the impurity scattering in σ bands. This makes the clean regime of σ bands particularly robust; in fact up to x=0.2 the upper critical fields can be well described within a clean limit model and its decrease is directly related to the suppression of the superconducting σ gap. For larger doping we observed the crossover to the dirty regime driven mostly by the strong decrease of $\Delta_\sigma$ rather than by the increase of $\Gamma_\sigma$.

This work is partially supported by I.N.F.M. through the PRA-UMBRA.

**Figure captions**

**Figure 1**: Left panel: $B_{c2}$ of the Al$_x$Mg$_{1-x}$B$_2$ samples as a function of temperature; right panel: $B_{c2}$ of the Al$_x$Mg$_{1-x}$B$_2$ samples as a function of the reduced temperature $T/T_c$.

**Figure 2**: $v_{F_{\sigma\,ab}}$ and $v_{F_{\sigma\,c}}$ as a function of x (upper panel); $\omega_{p_{\sigma\,ab}}$, $\omega_{p_{\sigma\,c}}$ and the spatial average $\omega_{p_\sigma}$ as a function of x (lower panel).

**Figure 3**: Main panel: $\Delta_\sigma$ (filled symbols) and $\Delta_\pi$ (open symbols) as a function of x obtained by Andreev reflection and specific heat data.[13] Inset: example of a point-contact conductance curve at T=4.2 K in the sample with x=0.2 (symbols). The line is a two-band BTK fit obtained with the following parameters: $\Delta_\sigma$=3.8 meV and $\Delta_\pi$=1.73 meV, the potential barrier parameters $Z_\sigma$ = 1.55 and $Z_\pi$ = 0.2 and the phenomenological broadening parameters $\gamma_\sigma$= 0.3 meV and $\gamma_\pi$ = 1.22 meV. The weight of the π-band conductance was taken equal to 0.75.

**Figure 4**: The rescaled field $B_{c2}^{res}(x)$ as a function of the reduced temperature $T/T_c$.

**Table captions**

**Table I.** Parameters of Al$_x$Mg$_{1-x}$B$_2$ polycrystalline samples: the critical temperature defined as $T_c$=(T$_{90\%}$ + T$_{10\%}$)/2 and $\Delta T_c$ =(T$_{90\%}$ - T$_{10\%}$), where T$_{90\%}$ and T$_{10\%}$ are estimated at the 90% and 10% of the resistive transition; $\rho_0$ defined as the resistivity measured at 40 K; the residual resistivity ratio defined as RRR=$\rho(300)/\rho(40)$; the crystallographic a and c axes; σ-band scattering rate $\Gamma_\sigma$=$\Gamma_{\sigma\sigma}$+$\Gamma_{\sigma\pi}$ given by the sum of intraband ($\Gamma_{\sigma\sigma}$) and interband ($\Gamma_{\sigma\pi}$) scattering rates.

**Table II.** Energy gaps and reduced energy gaps of Al$_x$Mg$_{1-x}$B$_2$ polycrystalline samples

**Table I**

| x | $T_c \pm \Delta T_c/2$ (K) | $\rho_0$ (μΩ·cm) | RRR | a axis (Å) | c axis (Å) | $\Gamma_\sigma$ (meV) | $\dfrac{\Gamma_\sigma}{\pi \Delta_\sigma}$ |
|---|---|---|---|---|---|---|---|
| 0    | 39.0±0.1 | 2.5±0.5 | 7.2 | 3.085 | 3.525 |    |     |
| 0.05 | 36.6±0.5 | 5.0±0.5 | 4.7 | 3.083 | 3.508 | 7  | 0.4 |
| 0.07 | 35.8±0.5 | 6.7±0.8 | 2.7 | 3.079 | 3.489 | 9  | 0.5 |
| 0.10 | 33.4±1.5 | 8±1     | 2.2 | 3.077 | 3.483 | 10 | 0.6 |
| 0.15 | 31.5±1.5 | 12±1    | 1.9 | 3.076 | 3.456 | 13 | 0.9 |
| 0.20 | 29.1±0.9 | 7±1     | 2.7 | 3.077 | 3.453 | 7  | 0.6 |
| 0.30 | 24±3     | 24±2    | 1.5 | 3.072 | 3.437 | 16 | 2.5 |

**Table II**

| x | $\Delta_\sigma$ (meV) | $\Delta_\pi$ (meV) | $2\Delta_\sigma/k_B T_c$ | $2\Delta_\pi/k_B T_c$ |
|---|---|---|---|---|
| 0    | 7.5±0.5 | 2.85±0.01 | 4.4±0.3 | 1.70±0.01 |
| 0.10 | 5.4±0.9 | 2.5±0.1   | 3.9±0.8 | 1.75±0.3  |
| 0.20 | 3.8±0.9 | 1.7±0.2   | 3.0±0.8 | 1.4±0.5   |
| 0.30 | 2.0±0.2 | 0.5±0.3   | 2.0±0.6 | 0.5±1     |

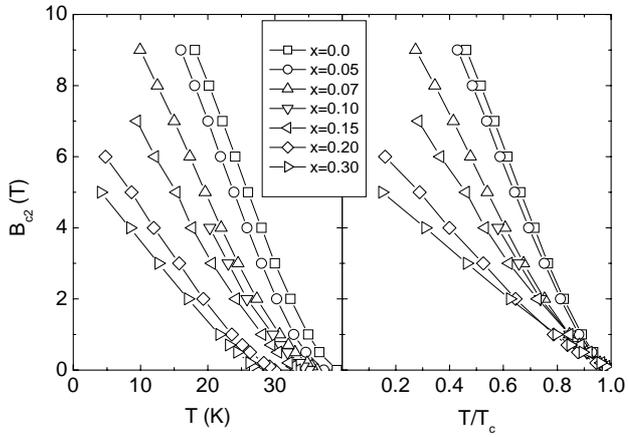

Figure 1

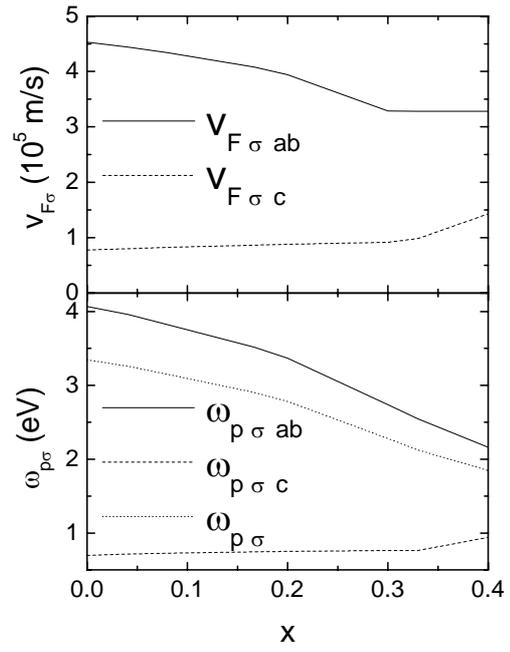

Figure 2

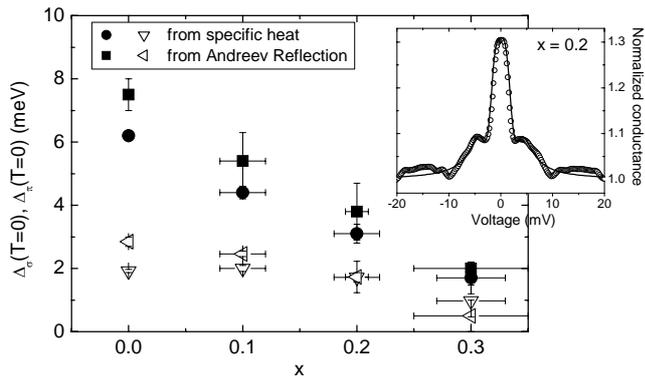

Figure 3

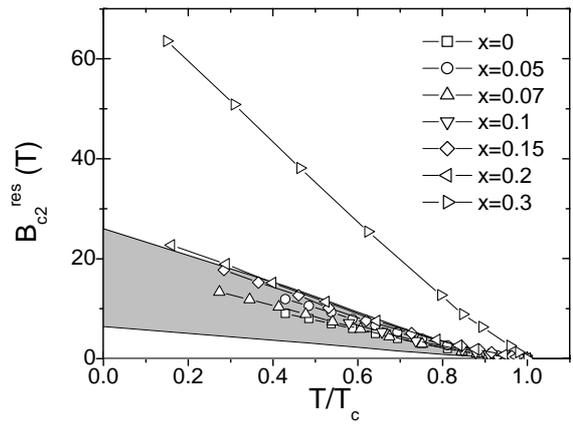

Figure 4

5